\documentclass[letterpaper]{article}
\usepackage{aaai18}
\usepackage{times}
\usepackage{helvet}
\usepackage{courier}
\usepackage{url}
\usepackage{graphicx}
\title{Ethical Considerations for AI Researchers}
\author{Kyle Dent \\ Palo Alto Research Center \\ kdent@parc.com }

\begin{document}
\maketitle
\begin{abstract}
Use of artificial intelligence is growing and expanding into applications that
impact people's lives. People trust their technology without really
understanding it or its limitations. There is the potential for harm and we are
already seeing examples of that in the world. AI researchers have an obligation
to consider the impact of intelligent applications they work on. While the
ethics of AI is not clear-cut, there are guidelines we can consider to minimize
the harm we might introduce.
\end{abstract}

\section{Introduction}
A quick scan of recent papers covering the area of AI and ethics reveals
researchers' admirable impulse to think about teaching intelligent agents human
values \cite{AAAIW1612582,AAAIW1612623,AAAIW1612624}. There is, however,
another important and more immediate aspect of AI and ethics we ought to take
into consideration. AI is being widely deployed for new applications; it's
becoming pervasive; and it's having an effect on people's lives. AI researchers
should reflect on their own personal responsibility with regard to the work
they do.  Many of us are motivated by the idea that we can contribute useful
new technology that has a positive impact on the world. Positive outcomes have
largely been the case with advanced technologies that improve cancer diagnosis
and provide safety features in cars, for example. With vast amounts of
computing power and a number of improved techniques, intelligent software is
being adopted in more and more contexts that affect people's lives. How people
use it is starting to matter, and the impact of our decisions matters.

Not surprisingly as the use of AI expands, negative consequences of its
failures and design flaws are more visible. Much of the AI that has recently
been deployed derives its intelligence from learning algorithms
that are based on statistical analysis of data. The acquisition, applicability,
and analysis of that data determine its output. Statistics shine when making
predictions about distributions over populations. That predictive power fades
when applied to individuals. There will be faulty predictions. The popular
press is rife with misuses of statistical analysis and AI
\cite{NYTIMES:AI,O'Neil:2016:WMD:3002861}. Given the growing use, the built-in
uncertainties, and the public's tendency to blindly trust technology, we have a
responsibility to consider the likely and unlikely outcomes of the choices we
make when we are designing and developing tools or predictive systems to
support decision making that affect people and communities of people.

Purposely malicious choices are obviously ethically unacceptable. In
\cite{DBLP:journals/corr/Yampolskiy15a}, the author outlines various pathways
that lead to dangerous artificial intelligence. Within the taxonomy, there are
pathways that introduce danger into artificial intelligence `on purpose.' The
other pathways inadvertently lead to hazards in the system. You can decide for
yourself if you are comfortable developing smart weapons, for example, and most
of us would, at a minimum, pause to consider the implications of that decision.
But the inadvertent pathways leading to dangerous AI can be difficult to
foresee and may come about from subtle interactions. Our less obvious
responsibility lies in giving careful consideration to our choices and being
clear to ourselves and our stakeholders about assumptions, trade-offs, and
choices we make.

Several other papers consider another ethical aspect in the fairness of
automatic systems
\cite{O'Neil:2016:WMD:3002861,DBLP:journals/corr/HardtPS16,NSTCCT}, and some
even conclude that it's inherently impossible for most problems
\cite{DBLP:journals/corr/KleinbergMR16}. One of the points I'll make is that
discussions about fairness and societal impact can be cut off once an
intelligent agent is introduced into the process. There is a popular feeling
that machines don't make value judgments and are inherently unbiased. However,
the assumptions we make when designing our systems are often based on
subjective value judgments; for example, choosing data sets, selecting
weighting schemes, balancing precision and recall. We have to be transparent
about what we do and be clear about the choices we have made. The ultimate
purpose matters and the decisions you come to must be communicated.

\section{Blind Trust in Technology}
Although there are pockets of skepticism towards intelligent systems, by and
large people are content to offload decisions to technology. In May 2016, there
was a widely publicized crash involving a Tesla Motors car being driven in
computer-assisted mode. It appears the driver had undue faith in the
capabilities of the car \cite{NHTSA:TESLA}. The following week another driver
following a GPS unit steered her car into Ontario's Georgian Bay
\cite{SUN:GPS}. These extreme examples reveal a trend in the general
population to trust the smart devices in our lives.

Ideally government agencies and jurisdictions would apply the principles of
open government and transparency when contracting with suppliers for
decision-making tools. In practice that hasn't been the case. Last year, two
researchers filed 42 open records requests in 23 different states asking for
information about software with predictive algorithms used by governments as
decision support tools \cite{YALE:TRANSPARENT}. Their goal was to understand
the policies built into the algorithms in order to evaluate their usefulness
and fairness. Only one of the jurisdictions was able to provide information
about the algorithms the software used and how it was developed. Some of those
who did not respond cited agreements with vendors preventing them from
revealing information, but many did not seem concerned about transparency in
their process nor the need to understand the technology. Assuming the best
intentions of the decision makers, they are also demonstrating great faith in
the technology and vendors they contract with.

There is also evidence that users of these systems, judges and hiring managers
for example, weight AI guidance too heavily. Without tools, when people are
making decisions, there is public awareness that decisions are made within some
context. We understand that individuals can be influenced even subconsciously
by their biases and prejudices. Technologically assisted decisions tend to shut
down the conversation about fairness despite their having a large effect on
people's lives. Those affected may not have the opportunity to contest the
decisions. If important decisions are made through our models, we must use care
in developing them and clearly communicate the assumptions we make.

\section{Ethical Obligations}
Physicians and attorneys have well-established codes of ethics. Doctors
famously commit to not doing any harm. Implied in that concept is the idea that
there is potential to do harm. It is clear from many examples, some of which I
mention in this paper, that there is the potential for harm in our work, and
given people's lack of understanding of the limits of and the trust they place
in technology, AI researchers have a personal, ethical obligation to reflect on
the decisions we make.

Ethical thinking helps us to make choices and just as importantly provides a
framework to reason about those choices. The framework we use (explicitly or
not) is defined by a set of principles that guide and support our decisions.
One of the difficult things about defining ethical standards is deciding the
values to base them on. Ethics issues will undoubtedly be discussed and argued
within the community and the world generally in the coming years. Each of us
can start by considering our own roles and being consciously aware of the
effects our work can have.

The stakeholders who decide to deploy intelligent decision making, government
agencies for example, generally aren't qualified to assess the assumptions,
models and algorithms in it. This asymmetrical relationship puts the burden on
those with the information to be clear, honest, and forthcoming with it. Those
at a disadvantage depend on us to inform them about technology's fitness for
their purpose, its reliability and accuracy. We usually focus on the technical
aspects of our work like selecting highly predictive models and minimizing
error functions, but when applying algorithmic decision-making that will
affect human beings, we have a responsibility to think about more.

\section{Recommendations for Consideration}
Ethics is not science. But it is possible to ground our thinking in
well-defined guidelines to assist in making ethical decisions for AI
development. A formal framework may even emerge within the researcher community
with time. In the short-term, the following is a list of thoughts and questions
to ask ourselves when designing predictive or decision-making systems.

\subsection{1. Relevance of data and models}
It is important to think carefully about the data used to train our technology.
Are the data and models appropriate to the real-life problem they are solving?
It is tempting to believe causal forces are at play when we find correlation on
a single dataset. Does the data capture the true variable of interest? Is it
consistent across observations and over time? We often introduce a proxy
variable because the variable we need isn't available or isn't easy to
quantify. Can your findings be calibrated against the real-world situation?
Even better could you measure the actual outcome you're trying to achieve?

In 2008, Researchers at Google had the idea that an increase in search queries
related to the flu and flu symptoms could be indicative of a spreading virus.
They created the Google Flu Trends (GFT) web service to track Google users'
search queries related to the flu. If they detected increased transmission
before the numbers from the U.S. Centers for Disease Control and Prevention
(CDC) came out, earlier interventions could reduce the impact of the virus. The
initial article reported 97\% accuracy using the CDC data as the gold standard
\cite{34503}. However, a follow-up report showed that in subsequent flu seasons
GFT predicted more than double what the CDC data showed \cite{154846}. Given
the first year's high accuracy, it would have been easy for the researchers to
believe they had discovered a strong, predictive signal. But online behavior
isn't necessarily a reflection of the real world. There are several factors
that might make the GFT data wrong. One of them is that the underlying
algorithms of Google Search itself (the GFT researchers don't control those)
can change from one year to the next. Also users' search behavior could have
changed. Mainly, however, people's search patterns are probably not a good
single indicator of a spreading virus. There are many other factors and various
reasons people might search for information.

Training data rarely aligns with real-life goals. In
\cite{DBLP:journals/corr/Lipton16a}, Lipton presents challenges to providing
and even defining interpretability of machine learning outputs. He identifies
several possible points of divergence between training data and real-life
situations. For example, off-line training data is not always representative of
the true environment, and real-world objectives can be difficult to encode as
simple value functions. Often we work with data that was collected for other
purposes and almost never under ideal, controlled circumstances. What was the
original purpose in collecting the data, and how did that determine its
content? In July of 2015, another group at Google had to apologize for its
Photos application identifying a black couple as gorillas
\cite{GUYNN:USATODAY}. Their training dataset was not representative of the
population it was meant to predict. Also, there are limits to the amount of
generalization we can expect from any learning method trained on a particular
dataset.

Is it possible your dataset contains biases? When making decisions related to
hiring, judicial proceedings, and job performance, for example, many personal
characteristics are legally excluded. Also, humans are good at discarding
variables they recognize as irrelevant to the decision to be made; computers
are blind to those considerations. Are there other characteristics that are
closely correlated with legally and ethically protected ones? If you don't
consider those, you can inadvertently treat people unfairly based on protected
or irrelevant characteristics. There is often a trade-off between accuracy and
the intelligibility of a model \cite{Caruana:2015:IMH:2783258.2788613}. More
predictive but harder-to-understand models can make it difficult to know which
personal characteristics determine the decision and are therefore not
available for validation against human judgment.

In \cite{Caruana:2015:IMH:2783258.2788613} the authors describe a system that
learned a rule that patients with a history of asthma have a lower risk of
dying from pneumonia. Based on the data used to train the system, their model
was absolutely correct. However, in reality asthma sufferers (without
treatment) have a higher risk of dying from pneumonia. Because of the increased
risk, when patients with a history of asthma go to the hospital, the general
practice is to place them in an intensive care unit. The extra attention they
receive decreases their risk of dying from pneumonia even below that of the
general population. It is our natural inclination to develop models with the
highest accuracy. However, the necessity of visibility into decisions where
people's lives are concerned, may increase the importance of explainability at
the expense of some predictive performance. In all cases, our stakeholders must
understand the decisions we make and the trade-offs implied by them.

\subsection{2. Safeguards for Failures and Misuse}
Even experienced researchers with the best intentions are inclined to favor the
positive outcomes of their work. We highlight positive results, but we should
also think through failure modes and possible unintended consequences. What
about misuse? There isn't a lot you can do about a person determined to use the
technology in ways it wasn't intended, but are there ways a good-faith user
might go wrong? Can you add protections for that?

The 2016 Tesla accident mentioned before was catastrophic. The driver used
computer-assisted mode in conditions it was expressly not designed for
resulting in his death.  The accident was investigated by two government
agencies. The first finding from the National Highway Traffic and Safety
Administration found that the driver-assist software had no safety defects and
declared that, in general, the vehicles performed as designed
\cite{NHTSA:TESLA} implying that responsibility for use of the system falls on
the operator. A later investigation from The National Transportation and Safety
Board found otherwise \cite{NTSB:TESLA}. They declared that the automatic
controls played a major role in the crash. The fact that the driver was able to
use computer assistance in a situation it was not intended for was problematic.
The combination of human error and insufficient safeguards resulted in an
accident that should not have happened.

\subsection{3. Accuracy}
How accurate is your algorithm and how accurate does it need to be? Do your
stakeholders understand the number of people who will be subject to a missed
prediction given your measure of accuracy? A model that misses only 1\% shows
phenomenally good performance, but if hundreds or thousands of people are still
adversely affected, that might not be acceptable. Are there human inputs that
can compensate for the system's misses and can you design for that? What about
post-deployment accuracy? Accuracy in training data doesn't always reflect real
usage. Do you have a way to measure runtime accuracy? The world is dynamic and
changes with time. Is there a way to continue to assess the accuracy after
release? How often does it have to be reviewed?

\subsection{4. Size and severity of impact}
Think about the numbers of people affected. Of course, you want to avoid
harming anyone but knowing the size or the severity of negative consequences
can justify the cost of extra scrutiny. You might also be able to design
methods that mitigate for them. Given an understanding of the impact, you can
make better decisions about the value required by the extra effort.

\section{Conclusion}

Individual researchers, especially in commercial operations, don't always have
the chance to communicate clearly and transparently with clients. At least
being transparent with your immediate stakeholders can set the right
expectations for them when they represent your work down the line.  You are
necessarily making decisions about the models and software you develop. If you
don't surface those decisions to discuss their effect, they may never be
brought to light.

A short paper cannot cover such a large and multi-faceted issue. The main idea
is for each of us to think individually about our own responsibilities and the
impact our work can have on real lives. It's useful to spend time thinking about
our assumptions and the trade-offs we make in the context of the people who
will be affected. Communicating those to everyone concerned is also critical.
Modern versions of the Hippocratic Oath are still used by many medical schools.
The spirit of the oath is applicable to most research affecting human beings.
One phrase is especially general and worth keeping in mind:

\begin{quote}
``I will remember that I remain a member of society, with special obligations
to all my fellow human beings\ldots'' \cite{TYSON:OATH}
\end{quote}

\bibliography{ethical_considerations}
\bibliographystyle{aaai}

\end{document}